\documentclass[prc,amsmath,amssymb,superscriptaddress,noffotinbib,twocolumn,preprintnumbers]{revtex4-1}
\usepackage{url,color,colordvi,epsfig,pstricks}
\usepackage[colorlinks=true, citecolor=blue, filecolor=blue, linkcolor=blue, urlcolor=blue, pdftex]{hyperref}
\usepackage{graphicx,bm,dcolumn}
\usepackage{hyperref,slashed,lineno}
\usepackage{physics}
\usepackage{inputenc}
\usepackage{mathtools}

\hypersetup{
    colorlinks=true, 
    linktoc=all, 
    linkcolor=blue, 
    citecolor=blue}

\newcommand{\Pv}{{\bf P}}
\newcommand{\qv}{{\bf q}}
\newcommand{\wfr}{\omega^{fr}}
\newcommand{\qfr}{{\bf q}^{fr}}

\newcommand{\diff}{\mathrm{d}}						 								
\newcommand{\sumint}{\mathclap{\displaystyle\int}\mathclap{\textstyle\sum}}
\begin{document}

\title{Relativistic effects in Green's function Monte Carlo calculations of neutrino-nucleus scattering}

\author{Alexis Nikolakopoulos}
\affiliation{Theoretical Physics Department, Fermi National Accelerator Laboratory, P.O. Box 500, Batavia, Illinois 60510, USA}

\author{Alessandro Lovato}
\affiliation{Physics Division, Argonne National Laboratory, Argonne, IL 60439}
\affiliation{Computational Science Division, Argonne National Laboratory, Argonne, IL 60439}
\affiliation{INFN-TIFPA Trento Institute for Fundamental Physics and Applications, 38123 Trento, Italy}

\author{Noemi Rocco}
\affiliation{Theoretical Physics Department, Fermi National Accelerator Laboratory, P.O. Box 500, Batavia, Illinois 60510, USA}

\date{\today}

\begin{abstract}
Microscopic calculations of neutrino-nucleus scattering cross sections are critical for the success of the neutrino-oscillation program. In addition to retaining nuclear correlations in the initial and final state of the reaction, they are based on consistent nuclear interactions and transition current operators, thereby enabling robust uncertainty quantification. In this work, we address a significant limitation of these microscopic methods, which arises from their nonrelativistic nature. By performing the calculations in a reference frame that minimizes nucleon momenta and utilizing the so-called ``two-fragment'' model, we extend the applicability of Green's function Monte Carlo calculations of neutrino-nucleus scattering to higher momenta than currently possible. To validate this approach, we compare our theoretical predictions against inclusive data measured by the MiniBooNE, T2K, and MINER$\nu$A experiments.

\end{abstract}

\maketitle

\preprint{FERMILAB-PUB-23-167-T}


\section{Introduction}

Over the past decade, there has been tremendous progress towards computing lepton-nucleus scattering within the so-called {\it ab initio} nuclear many-body approaches, which provide a microscopic description of nuclear dynamics starting from the individual interaction among the constituent neutrons and protons~\cite{Hergert:2020bxy}. By exploiting integral-transform techniques, both quantum Monte Carlo (QMC)~\cite{Carlson:2014vla,Lynn:2019rdt,Gandolfi:2020pbj} and coupled-cluster~\cite{Hagen:2013nca} methods retain nuclear correlations in both the initial and final states of the scattering process. Notably, the latter are generated by realistic nuclear Hamiltonians consistent with the electroweak-current operators entering the transition matrix element. 
Hence, in addition to providing an accurate description of nuclear dynamics, microscopic approaches allow one to estimate the theoretical uncertainties associated with modeling nuclear dynamics. This aspect is particularly relevant for the accelerator-neutrino program, as cross-section uncertainties represent a significant component of the error budget of neutrino-oscillation parameters~\cite{Benhar:2015wva,Katori:2016yel,Alvarez-Ruso:2017oui}. 

Among QMC methods, Green's function Monte Carlo (GFMC) has been extensively employed to compute the electroweak response functions of nuclei with up to $A=12$ nucleons starting from imaginary-time propagators, corresponding to their Laplace transforms. GFMC calculations of inclusive electron and neutrino cross sections of $^4$He and $^{12}$C are in excellent agreement with experimental data~\cite{Lovato:2016gkq,Lovato:2017cux,Lovato:2020kba}. CC can reach larger systems due to its favorable polynomial scaling with the number of nucleons. After its initial application to low-energy nuclear dipole responses~\cite{Bacca:2013dma}, the coupled-cluster approach was extended to compute the Coulomb sum rule of $^{4}$He and $^{16}$O~\cite{Sobczyk:2020qtw}. Most recently, the Authors of Ref.~\cite{Sobczyk:2021dwm} have carried out coupled-cluster calculations of the longitudinal electromagnetic response function of $^{40}$Ca, finding very good agreement with experiments in the quasi-elastic region. 

One of the main limitations of both QMC and coupled-cluster approaches has to be ascribed to the nonrelativistic formulation of the many-body problem. Although the leading relativistic corrections are typically included in the transition operators~\cite{Shen:2012xz}, the kinematics of the reaction is nonrelativistic, thereby limiting the application of these methods to moderate values of the momentum transfer. This restriction is particularly relevant when making predictions for inclusive neutrino-nucleus cross sections since the incoming neutrino flux is not monochromatic. Its tails extend to energies where relativistic effects cannot be neglected. 

In a number of works~\cite{Efros:2005uk,Efros:2009qp,Yuan:2010gh,Efros:2010fe,Yuan:2011rd,Rocco:2018}, a method was proposed to
extend the applicability of manifestly nonrelativistic hyperspherical-harmonics and QMC methods to higher momentum transfer values than typically possible. This method reduces relativistic effects by performing the calculations in a reference frame that minimizes nucleon momenta. Additional relativistic effects in the kinematics are accounted for by employing the so-called two-fragment model, which allows one to obtain, in a relativistically-correct way, the kinematic inputs of the nonrelativistic dynamical calculations. 

In this work, we quantify the role of relativistic effects in the GFMC calculations of the electroweak response function of $^{12}$C induced by charged-current transitions by analyzing their frame dependence with and without the two-fragment model. Following the strategy discussed in Ref.~\cite{Rocco:2018}, we compute inclusive neutrino-$^{12}$C scattering cross sections choosing a reference frame that minimizes these effects. We compare our theoretical calculations with experimental data measured by the MiniBooNE~\cite{Mini}, T2K~\cite{T2K}, and MINER$\nu$A~\cite{MINERVA} experiments. Note that their neutrino fluxes are characterized by different energy distributions, whose high-energy tails extend beyond the GeV region. 

This paper is organized as follows. In Section~\ref{sec:relativistic_effects}, we outline the connection between inclusive neutrino-nucleus cross sections and electroweak response functions, review the Lorentz transformations to different reference
frames, and apply them to the GFMC electroweak response functions. In Section~\ref{sec:results_responses}, we gauge the role of relativistic effects in the charged-changing response functions, while inclusive cross-section results are discussed in Section~\ref{sec:cross_sections}. Finally, in Section~\ref{sec:conclusions}, we draw our conclusions and outline future perspectives of this work.

\section{Implications of relativity for nuclear responses}
\label{sec:relativistic_effects}

\subsection{Nuclear responses and charged-current cross section}
The differential cross section for inclusive charged-current (CC) scattering of a neutrino with the nucleus can be written as
\begin{align}
    \frac{\diff \sigma}{\diff E_l \diff \Omega_l} &= \frac{G^2}{4\pi^2} k_l E_l ( v_{CC} R_{CC} - v_{CL} R_{CL} + v_{LL}R_{LL}  \nonumber \\
    \label{eq:CS}
     &+ v_{T} R_{T} + v_{T^\prime} R_T{^\prime} ),
\end{align}
with $G = G_F\cos\theta_c$, and $E_l$, $k_l$ denote the energy and momentum of the final-state lepton, respectively.
The decomposition into factors $v_X$ that depend only on the lepton kinematics, and nuclear responses $R_X$ follows from considering a single boson exchange.
The expressions for the lepton factors can be found in Refs.~\cite{Lovato:PRX}.
The inclusive nuclear electroweak response functions correspond to specific elements of the hadron tensor, defined as
\begin{align}
R^{\mu\nu} &=\sum_f \langle \Psi_0 \rvert {J^\mu}^\dagger(\omega,\mathbf{q}) \lvert \Psi_f \rangle \langle\Psi_f \rvert J^\nu(\omega,\mathbf{q}) \lvert \Psi_0 \rangle\nonumber\\
&\times \delta\left(\omega + E_0 - E_f\right),
\end{align}
where $|\Psi_0\rangle$ and $|\Psi_f\rangle$ denote the nuclear initial ground-state, and final bound- or scattering-state of energies $E_0$ and $E_f$. The nuclear electroweak current $J^\mu(\omega,\mathbf{q})$ depends upon the energy and momentum transferred to the nuclear system $\omega = E_\nu - E_l$, and $\qv = \mathbf{k}_\nu - \mathbf{k}_l$. Without loss of generality, we take $\qv$ to be parallel to the $z$-axis, so that the five inclusive nuclear responses in Eq.~(\ref{eq:CS}) can be expressed as
\begin{align}
R_{CC}(\omega,q) &= R^{00}     (\omega,q)     , \nonumber\\
R_{CL}(\omega,q) &= 2\Re R^{0z}(\omega,q)     , \nonumber\\
R_{LL}(\omega,q) &= R^{zz}     (\omega,q)     , \nonumber\\
R_{T} (\omega,q) &= \frac{R^{xx} + R^{yy}}{2}(\omega,q) , \nonumber\\
R_{T^\prime}(\omega,q) &= 2\Im R^{xy} (\omega, q)\, ,
\end{align} 
where $q = \lvert \qv \rvert$. 
The longitudinal contribution to the cross section can be written to make the dependence on lepton mass explicit as
\begin{align}
\label{eq:massterms}
    &v_{CC}R_{CC} - v_{CL}R_{CL} + v_{LL}R_{LL} = \nonumber  \\ 
    &v_{CC} R_{L} - \frac{m_l^2}{qE_l} R_{CL} + \frac{m_l^{2}}{q^2}\left[2\frac{E_\nu}{E_l} - v_{CC} \right] R_{LL}.
\end{align}
Hence, the following combination of response functions
\begin{equation}
    \label{eq:RL}
    R_L \equiv R_{CC} - \frac{\omega}{q} R_{CL} + \left(\frac{\omega}{q} \right)^2 R_{LL},
\end{equation}
yields the leading longitudinal contribution when the momentum transfer and lepton energy are large compared to the outgoing lepton mass.

\subsection{Lorentz transformations to different reference frames}
The laboratory frame (LAB) is the reference frame in which the initial nucleus is at rest,  $\Pv_i = 0$. In this work, we evaluate the electroweak response functions in different reference frames which move with respect to the LAB frame along the direction specified by the momentum transfer $\qv$. 

Since the inclusive electroweak currents transform as four-vectors under a Lorentz-boost, the hadron tensor elements transform as
\begin{equation}
R^{\mu\nu}_{LAB}(\omega,q) = B^{\mu}_{~\alpha}\left[\beta\right] B^{\nu}_{~\beta} \left[\beta\right] R_{fr}^{\alpha\beta}(\wfr,\qfr)\,.
\end{equation}
In the last equation, $B$ indicates a Lorentz boost, and $R_{fr}$ the response evaluated in a frame that moves with relative velocity $\boldsymbol{\beta}$ with respect to the LAB frame.
For boosts along $\qv$, one can write $B$ in matrix notation as
\begin{equation}
\label{eq:boostmatrix}
B^{\mu}_{~\nu} = 
\begin{pmatrix}
\gamma & 0 & 0 & \gamma \beta \\
0 & 1 & 0 & 0 \\
0 & 0 & 1 & 0 \\
\gamma\beta & 0 & 0 & \gamma
\end{pmatrix} ,
\end{equation}
where $\beta=|\boldsymbol{\beta}|$ and $\gamma = 1/\sqrt{1-\beta^2}$.
Whilst the transverse responses are unchanged by a boost along $\qv$, the longitudinal responses transform as 
\begin{align}
\label{eq:RCC_boost}
R_{CC}^{LAB} =& \gamma^2 \left[ R^{fr}_{CC} +  \beta^2 R^{fr}_{LL} + \beta R_{CL}^{fr} \right] \\
\label{eq:RLL_boost}
R_{LL}^{LAB} =& \gamma^2 \left[ R^{fr}_{LL} +  \beta^2 R^{fr}_{CC} +\beta R_{CL}^{fr} \right]\\
\label{eq:RCL_boost}
R_{CL}^{LAB} =& \gamma^2 \left[ 2\beta\left(R_{CC}^{fr} + R_{LL}^{fr} \right) + (1+\beta^2) R_{CL}^{fr}\right]\,.
\end{align} 

The energy and momentum transfer in the moving frame are connected to the ones in the LAB frame by the inverse boost
\begin{align}
   \qfr = \gamma(\qv  - \boldsymbol{\beta} \omega), \quad \wfr = \gamma(\omega -\beta q),
   \label{eq:inverse_boost}
\end{align}
thus one can write the boost parameter as
\begin{equation}
\label{eq:gamma_wq}
\gamma = \frac{\omega q + \wfr q^{fr}}{\omega q^{fr} + \wfr q}\, ,
\end{equation}
where $q^{fr}=|\qfr|$. 

When the nuclear current is conserved, as in the electromagnetic case and the vector contribution to the electroweak current, one has $\omega J^{0}(\omega,\qv) - \qv \cdot \mathbf{J}(\omega,\qv) = 0$, which implies 
\begin{align}
\label{eq:CVC_resp}
R_{CC}^{fr}(\omega^{fr},\qfr) &= \frac{q^{fr}}{2\omega^{fr}}R_{CL}^{fr}(\omega^{fr},\qv^{fr}) \nonumber\\ 
&= \left(\frac{q^{fr}}{\omega^{fr}}\right)^2 R_{LL}^{fr}(\omega^{fr}, \qv^{fr})\,.
\end{align}
Substituting the above relation in Eq.~(\ref{eq:RCC_boost}), and using Eq.~\eqref{eq:gamma_wq} one finds that
\begin{equation}
\label{eq:boost_wCVC}
R_{CC}(\omega,\qv) = \left(\frac{q}{q^{fr}}\right)^2 R_{CC}^{fr}\left(\wfr, \qfr\right).
\end{equation}
This is the relation used in Refs.~\cite{Rocco:2018} for electromagnetic interactions.
However, in this work, we consider electroweak transitions in which the axial contribution is not conserved. Therefore, we use the more general expressions for the Lorentz-boosts between different frames of Eqs.~(\ref{eq:RCC_boost} - \ref{eq:RCL_boost}).
Following Refs.~\cite{Efros:2005uk}, we introduce a phase-space factor to account for the covariant normalization of the initial target state; the full result for LAB frame responses reads
\begin{equation}
\label{eq:Boost_final}
R_{LAB}^{\mu\nu}(\omega,\qv) = \frac{E_i^{fr}}{M_A} B^\mu_{~\alpha} B^{\nu}_{~\beta}~ R^{\alpha\beta}_{fr}(\omega^{fr},\qv^{fr}),
\end{equation}
with $E_i^{fr} = \sqrt{(\Pv_i^{fr})^2 + M_A^2}$ where $M_A$ is mass of the nucleus.

Following Ref.~\cite{beck1992relativistic}, we introduce the active-nucleon $\zeta$-frame as the one in which $\Pv_i^{fr} = -(1-\zeta) A \qv^{fr}$, where, clearly, $A=12$ for $^{12}$C.  
The Lorentz boost that connects these momenta to the LAB frame energy reads
\begin{equation}
    (1-\zeta)\qv^{fr} = -\frac{\Pv_i^{fr}}{A}  = \boldsymbol{\beta}\gamma\frac{M_A}{A}.
\end{equation}
Using the inverse boost expression for $\qfr$ found in Eq.~\eqref{eq:inverse_boost}, the relative velocity reads
\begin{equation}
    \boldsymbol{\beta} = \frac{(1-\zeta)\qv}{M_A/A + (1-\zeta)\omega}.
\end{equation}
For $\zeta=1$ we recover the LAB frame, while different values $\zeta$ parameterize other reference frames. In particular $\zeta=1/2$ corresponds to the active nucleon Breit (ANB) frame with $\Pv_i = -A \qv^{ANB}/2$. 
In the vicinity of the quasielastic peak, the momentum is mostly absorbed by a single ``active'' nucleon, with a momentum of approximately $\mathbf{p}_i^{A} = \Pv_i^{fr}/A$.
For the active-nucleon $\zeta$-frames, we thus have
\begin{equation}
    \mathbf{p}_i^{A} = (\zeta-1)\qv^{fr},\quad \mathbf{p}_f^{A} = \zeta\qv^{fr}
\end{equation} 
Hence, in the ANB the magnitude of the active nucleon momentum in initial and final state is minimal. 
Moreover, the energy transfer at the quasielastic peak in the ANB frame is zero and this holds true for both the relativistic and nonrelativistic case implying that the responses peak in the same position. As a consequence, $\qv^{fr}$ at the quasielastic peak is also minimal in the ANB. For these reasons, the ANB frame has been chosen in Refs.~\cite{Efros:2005uk,Efros:2009qp,Rocco:2018} as the one that minimizes the effect of relativistic corrections to the kinematics. 

In addition to the LAB and ANB frames, in Ref.~\cite{Efros:2005uk,Efros:2009qp,Rocco:2018}, the electromagnetic response functions of nuclei with A=3, 4 are also evaluated in the anti-lab frame, defined by $\Pv_i = -\qv^{AL}$, and in the Breit frame with $\Pv_i = -\qv^{B}/2$. However, in the limit of large $A$ these frames tend to become indistinguishable from the LAB frame. For this reason, since we are considering a heavier target than those studies in Refs.~\cite{Efros:2005uk,Efros:2009qp,Rocco:2018}, we will only focus on the active-nucleon frames in which the momentum of the nucleus scales with $A$.

\subsection{Nuclear responses in different reference frames}

Within the GFMC the responses are computed in the ``intrinsic system'', in which the total center of mass motion of the nuclear system is zero, e.g.
\begin{align}
R^{int}(\omega^\prime,\qv^\prime)= &
\sumint~~\delta\left(\omega^\prime + \epsilon_0 - \epsilon_f\right) \nonumber\\
\times & \langle \Psi_0 \rvert {J^\mu}^\dagger \lvert \Psi_f \rangle \langle\Psi_f \rvert J^\nu \lvert \Psi_0 \rangle\,, 
\end{align}
where $\epsilon_0$ and $\epsilon_f$ are the intrinsic energies of the initial and final states, respectively, which are assumed to be frame independent.
The nonrelativistic response in the LAB frame can be recovered by setting $\qv^\prime = \qv$ and $\omega^\prime= w - \qv^2/(2M_A)$. While for a generic reference frame, the response functions can be obtained by identifying $\qv^\prime$ with the boosted momentum transfer
\begin{equation}
\qv^\prime = \qfr = \gamma\left(\qv -\beta \omega \right).
\end{equation}
and including the center of mass energies of the initial and final states in the energy transfer definition as $\omega^\prime= \omega^{fr}-(\Pv^{fr}_f)^2/(2M_A)+(\Pv^{fr}_i)^2/(2M_A)$. 

Note that the ``intrinsic system'' cannot be interpreted as a reference frame. Hence, $R_{CC}$ and $R_{LL}$ can be recovered from the single response $R_{LL}$ using current conservation as in Eq.~\eqref{eq:CVC_resp} only after the nonrelativistic $R_{LL}$ in a given frame is computed.

\begin{figure*}
    \includegraphics[width=0.325\textwidth]{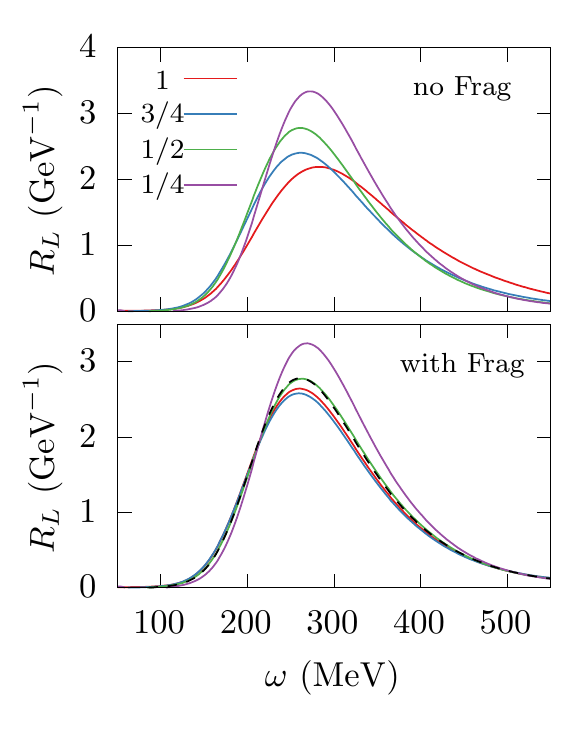}
    \includegraphics[width=0.325\textwidth]{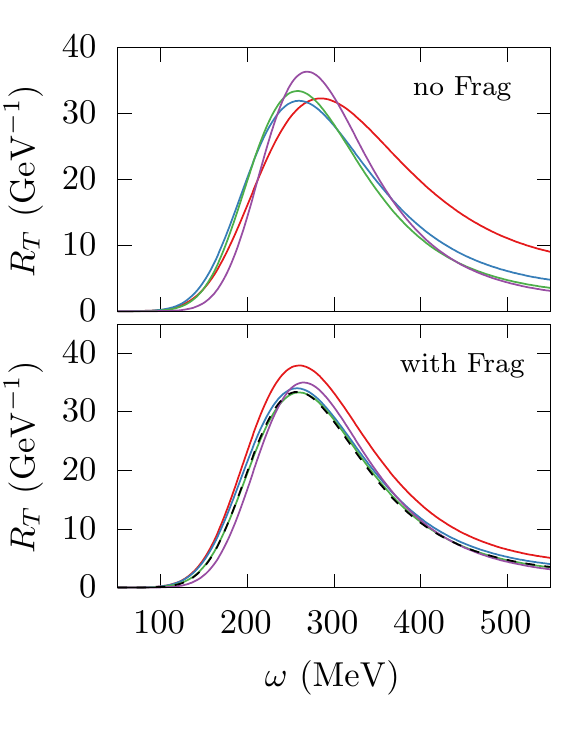}
    \includegraphics[width=0.325\textwidth]{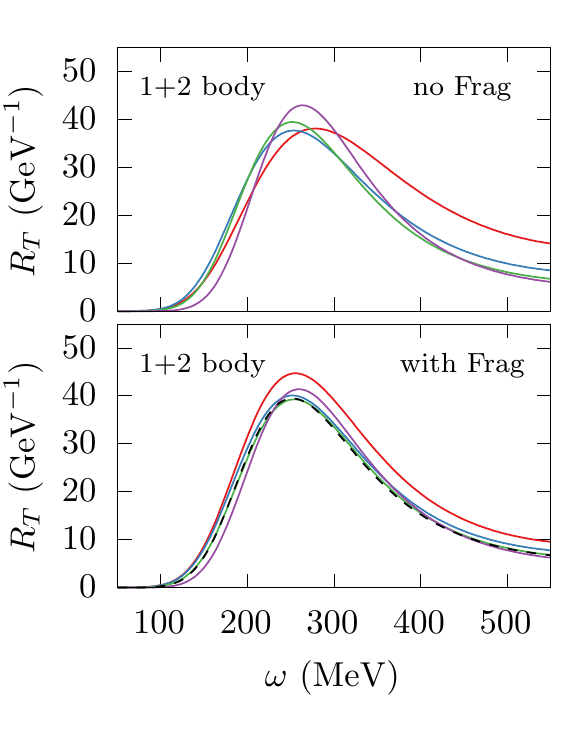}
    \caption{ Longitudinal and transverse electroweak response functions of $^{12}$C for different active-nucleon frames, parametrized by different values of $\zeta = 1,~3/4,~1/2,~1/4$. The top panels do not use the two-fragment model while the bottom panels do. The dashed black line in the bottom panels corresponds to the ANB result which does not include the fragment model.}
    \label{fig:C12_GFMC_w_wofrag}
\end{figure*}

\subsubsection{Two Fragment model}
The kinematics of quasielastic processes can be modified to account for relativistic corrections by employing the two-fragment model introduced in Ref.~\cite{Efros:2005uk}. This approach assumes that the dominant reaction mechanism in the quasielastic region is the break-up of the nucleus into two fragments, namely a knocked-out nucleon and a remnant $(A-1)$ system. Under this assumption, the energy of the hadronic final system can be written in a relativistically correct fashion as 
\begin{align}
\label{rel_kin}\nonumber
E_f^{fr} &= \sqrt{m^2 + ({\bf p}^{\rm fr}_f + (\mu/M_{A-1}){\bf P}_f^{\rm fr})^2} \\
         &+  \sqrt{M_{A-1}^2 + ({\bf p}^{\rm fr}_f - (\mu/m){\bf P}_f^{\rm fr})^2} \,;
\end{align}
where $\mu = \frac{m M_{A-1}}{m + M_{A-1}}$ is the reduced mass, ${\bf P}_f^{\rm fr}$ and ${\bf p}^{\rm fr}$ are the center of mass and relative momentum, respectively. Following the arguments of Ref.~\cite{Efros:2005uk,Rocco:2018}, we assume that both ${\bf P}_f^{\rm fr}$ and ${\bf p}^{\rm fr}$ are directed along ${\bf q}^{\rm fr}$. The value of ${\bf p}^{\rm fr}$ can be obtained by solving this equation and it has to be replaced in the definition of the intrinsic energy 
\begin{align}
    \label{eq:eps_2frag}
    \epsilon_f= \frac{({\bf p}_f^{\rm fr})^2}{2\mu}+\epsilon_0^{A-1}
\end{align}
where $\epsilon_0^{A-1}$ is the energy of the remnant nucleus. A detailed discussion on how to rewrite the energy conserving $\delta$ as a function of $\epsilon_f$ can be found in Ref.~\cite{Efros:2005uk}. 

\section{Results for Transformed responses}
\label{sec:results_responses}
Figure~\ref{fig:C12_GFMC_w_wofrag} shows the CC electroweak response functions of $^{12}$C computed in different active nucleon $\zeta$-frames and boosted back to the LAB fram applying the Lorentz transformation of Eq.~\eqref{eq:Boost_final}. The left panels display the longitudinal responses defined in Eq.~\eqref{eq:RL}. For momentum transfers where relativistic effects become important, the mass terms in Eq.~(\ref{eq:massterms}) are negligible even for muon-neutrino interactions, and $R_L$ determines the longitudinal cross section.

The results obtained in this work are consistent with those reported in Ref.~\cite{Rocco:2018}, which focused on the electromagnetic response functions of $^4$He. The two-fragment model is suitable to mitigate most of the frame-dependence in the nonrelativistic calculations, as the responses computed in different frames collapse onto a single curve. This behavior has to be confronted with the top panels, in which the the two fragment model is not applied. There, a significant frame dependence is visible, in both the logitudinal and transverse channel. As expected, the longitudinal and transverse CC responses obtained in the ANB-frame ($\zeta = 1/2$) are largely unaffected by the use of the two fragment model. To better appreciate this behavior, the dashed black line in the bottom panels of Fig.~\ref{fig:C12_GFMC_w_wofrag} corresponds to the results obtained in the ANB frame without employing the two-fragment model.

As shown in the rightmost panel of Fig.~\ref{fig:C12_GFMC_w_wofrag}, displaying the CC transverse response functions, the same behavior persists even when two-body current contributions are significant. Similarly to the one-body case, we observe that applying the two-fragment model to the total transverse response reduces the frame dependence of the calculation, with all curves aligning on the ANB frame one. Hence, we can infer that the single-nucleon knockout remains the dominant reaction mechanism even when two-body contributions are included in the current operator. This finding is is consistent with Refs.~\cite{Fabrocini:1997,Franco-Munoz:2022jcl}, 
whose Authors argue that the transverse enhancement is primarily due to the constructive interference between one- and two-body currents, leading to single-nucleon knockout final states. Notably, these works are based on completely different models of nuclear dynamics, namely the correlated basis function theory and the relativistic mean field. 
\begin{figure}
    \centering
    \includegraphics[width=0.49\textwidth]{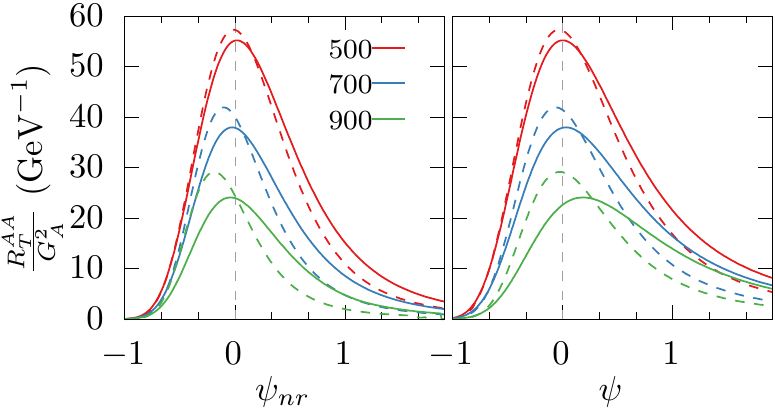}
    \caption{Energy transfer dependence of the transverse response in the LAB (solid) and ANB (dashed) frames with the two-fragment model for different values of $q$. The left (right) panel shows the responses as a function of the nonrelativistic (relativistic) scaling variables.}
    \label{fig:C12_scaling_r_nr}
\end{figure}

The results obtained in the ANB (with or without the two fragment model), incorporate relativistic corrections to the kinematics. This is shown explicitly in Fig.~\ref{fig:C12_scaling_r_nr}, where we compare the energy-dependence of the response in the LAB and ANB frames, for different values of $q$ as a function of the relativistic (left panel) and nonrelativistic (right panel) scaling variable~\cite{Alberico:1988, Rocco:2017}, which is defined as
\begin{equation}
    \psi_{nr}(\omega,q) = \frac{m}{|\qv| k_F}\left(\omega - \frac{\qv^2}{2m} - \epsilon_{nr} \right)
\end{equation}
where the Fermi momentum for $^{12}$C is taken to be $k_F=225$ MeV, and the energy shift $\epsilon_{nr} \sim 40$ MeV is included to center the peaks at $\psi_{nr}=0$. 
It is clear that the LAB results, corresponding to the solid lines, exhibit a universal energy-dependence in terms of $\psi_{nr}$ for the three different values of momentum transfer: $q=500$, $700$, and $900$ MeV. On the other hand, the peaks of the responses obtained using the two-fragment model (or the ANB) are shifted to smaller $\psi_{nr}$, while the high-$\psi_{nr}$ tail shrinks more rapidly, as $q$ increases. The same responses are shown in the right-hand panel, as function of the relativistic scaling variable~\cite{Alberico:1988,Barbaro:1998gu} 
\begin{equation}
    \psi(\omega,q) = \frac{1}{\xi_F} \frac{\lambda^{\prime} - \tau}{\left[ \tau(1+\lambda^{\prime}) + \kappa \sqrt{\tau(\tau+1)} \right]^{1/2}},
\end{equation}
with the dimensionless variables defined as
\begin{align}
    \lambda^\prime &= \frac{\omega - \epsilon_r}{2M_N},~\kappa = \frac{|\qv|}{2M_N},~\tau = \frac{Q^2}{4M_N^2} \\ 
    \xi_F &= \sqrt{1+\left(\frac{k_F}{M_N}\right)^2} - 1.
\end{align}
In the definition of $\psi$ we set $\epsilon_{r} \sim 30~\mathrm{MeV}$ so as to aligh the peak of the ANB responses at approximately $\psi=0$. Comparing the different dashed lines, it emerges that the ANB results are aligned when plotted as a function of the relativistic scaling variable, thus confirming that relativistic effects are properly accounted for in the ANB frame. On the other hand, the nonrelativistic responses evaluated in the LAB frame manifestly violate relativistic scaling.

\begin{figure}[b]
    \includegraphics[width=0.97\columnwidth]{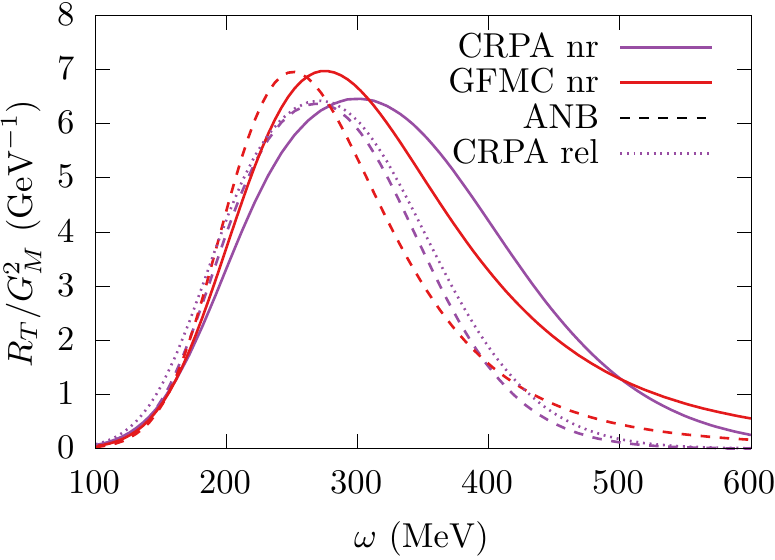}
    \caption{CC vector transverse response functions at $q=700~\mathrm{MeV}$.  The red and purple curves display the GFMC and CRPA results. The solid lines show the fully nonrelativistic calculations while the dashed ones have been obtained computing the response in the ANB frame. The dotted lines implement the shift of outgoing nucleon energies (see Eq.~\eqref{eq:TNshift}).}
    \label{fig:Frag_vs_shift}
\end{figure}

\begin{figure*}
    \centering
    \includegraphics[width=0.3\textwidth]{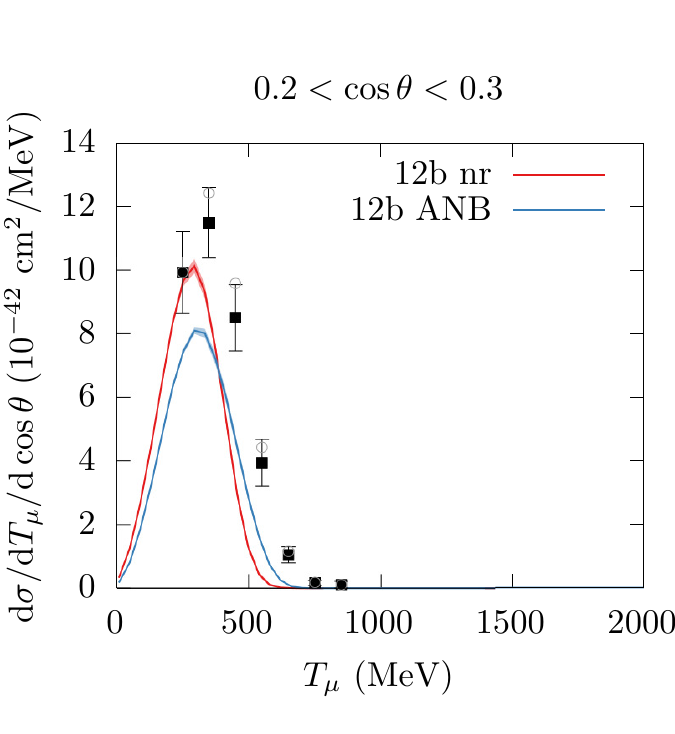}
    \includegraphics[width=0.3\textwidth]{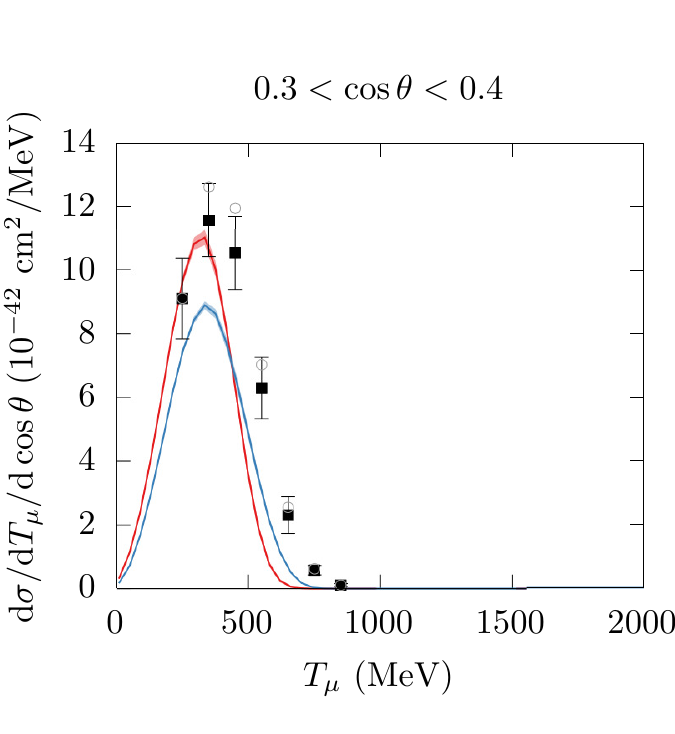} 
    \includegraphics[width=0.3\textwidth]{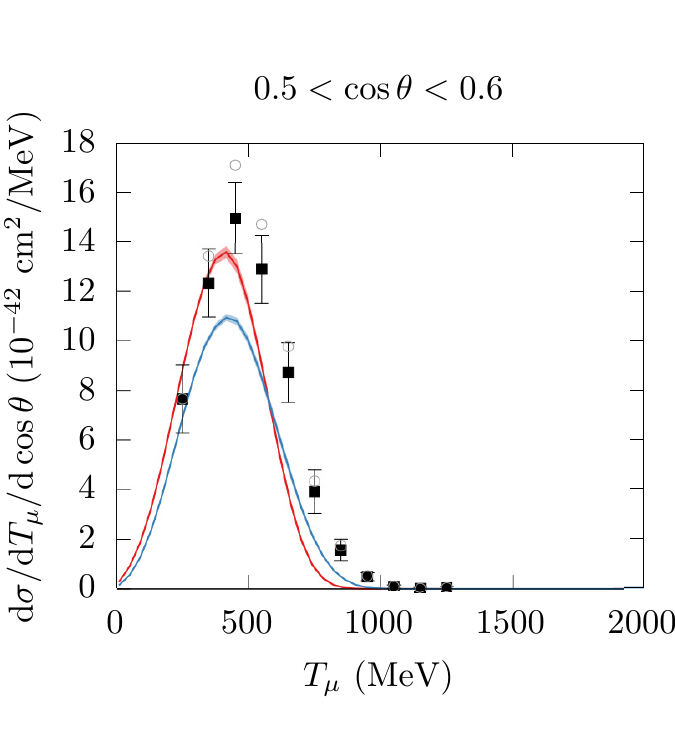} \\
    \includegraphics[width=0.3\textwidth]{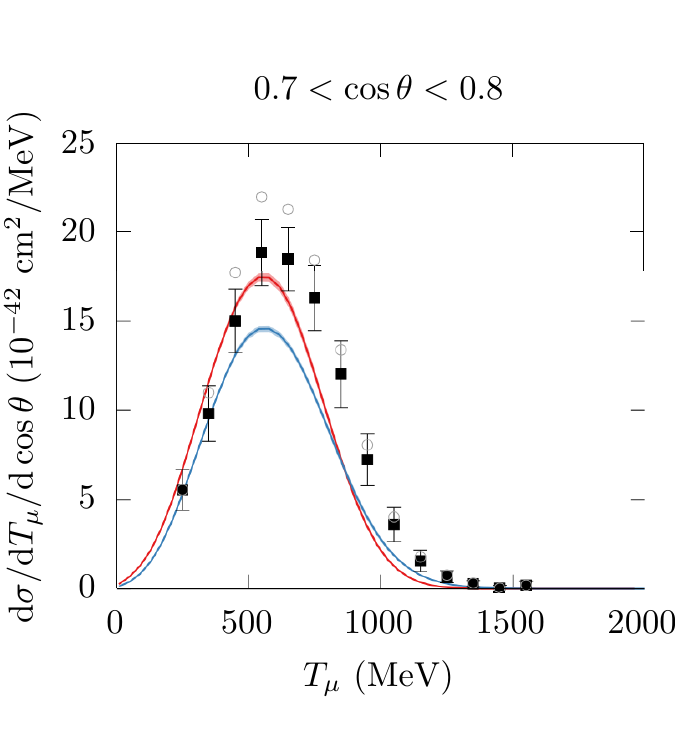}
    \includegraphics[width=0.3\textwidth]{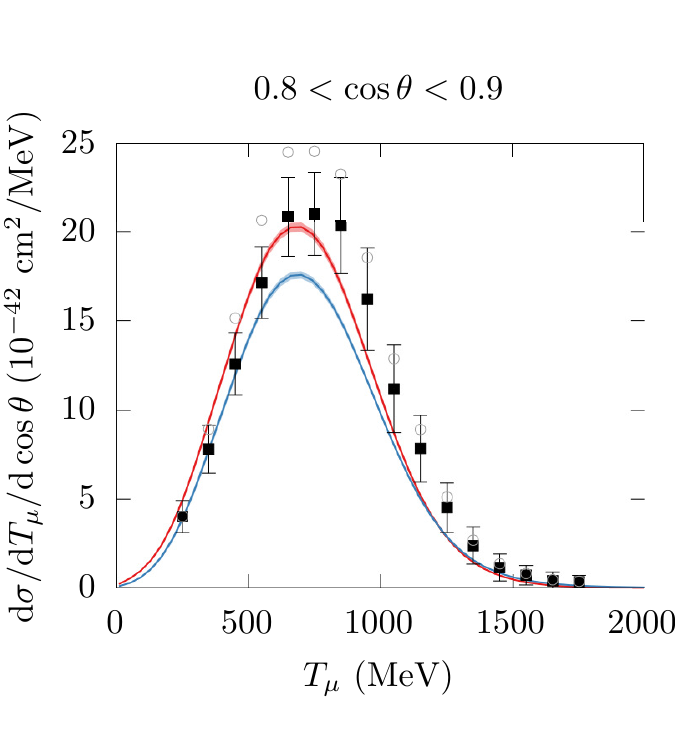} 
    \includegraphics[width=0.3\textwidth]{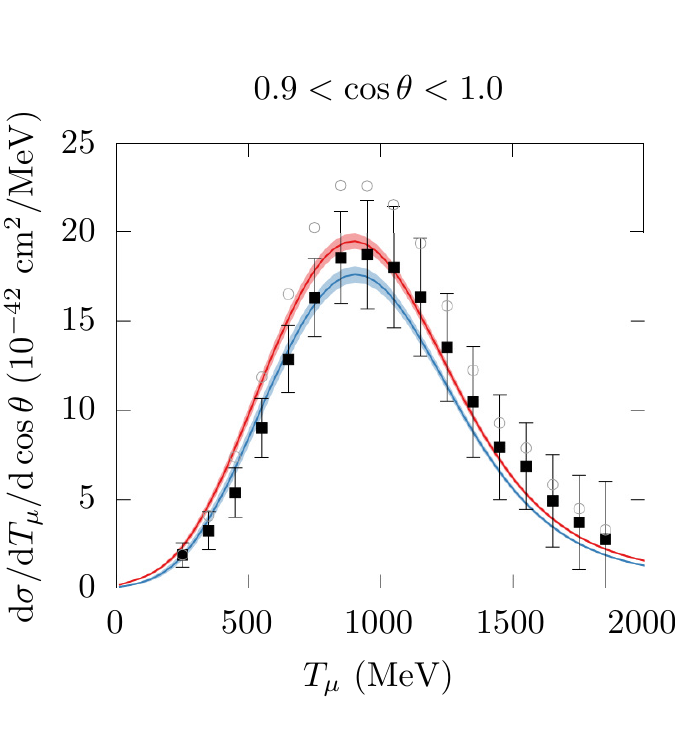}
    \caption{Flux averaged double differential cross section for MiniBooNE. The nonrelativistic GFMC results (nr) are compared to the results obtained in the ANB. They both include one- and two-body current contributions. The open circles are the cross section to which the background reported in Ref.~\cite{MB:QECS} is added.}
    \label{fig:MB_fluxfolded_12b}
\end{figure*}

For benchmark purposes, we consider alternative schemes that have been develop to account for relativistic effects in nonrelativistic calculations. In Refs.~\cite{Amaro:2005}, relativistic corrections for nucleon knockout in a nonrelativistic shell model are implemented by shifting the outgoing nucleon energy when solving the Shr\"{o}dinger equation as
\begin{equation}
\label{eq:TNshift}
    T_N \rightarrow T_{N}^\prime = T_N\left(1 + \frac{T_N}{2m} \right)\, .
\end{equation}
Since the the nonrelativistic kinetic energy is $p^2 = 2m T_N^\prime$, the above shift corresponds to using the relativistic momentum $p^2 = T_N(2m + T_N)$, thereby effectively transforming the nonrelativistic Shr\"{o}dinger equation into a form similar to a radial Dirac equation for the upper components of the spinors~\cite{PhysRevC.22.2027}. The latter indeed uses as ``energy'' $p^2/(2m)$, $p$ being the relativistic momentum.
The effect of this substitution in a CRPA calculation of the transverse response~\cite{Jachowicz:EPST,Jachowicz02} is shown in Fig.~\ref{fig:Frag_vs_shift}. Note that in Ref.~\cite{Pandey:2016}, the CRPA results additionally includes the leading order correction to the electroweak currents of Ref.~\cite{Jeschonnek}. In this figure, we compare the effect of shifting the kinetic energy of the nucleon as in Eq.~\eqref{eq:TNshift} with computing the response in the ANB fram and then boosting it back to the LAB frame. It is clear that both approaches lead to very similar $\omega$ dependence of the corrected responses. Note that the shift of Eq.~(\ref{eq:TNshift}) cannot be readily implemented to correct the GFMC responses. However, comparing with Eq.~\eqref{eq:eps_2frag}, the shift of Eq.~\eqref{eq:TNshift} resembles applying the two-fragment model in the LAB frame in the limit of large $A$, i.e. using the kinetic energy derived from the relativistic momentum as discussed above.

\section{Flux-Averaged cross sections}
\label{sec:cross_sections}
We compute the CC inclusive cross sections for different kinematic setups, relevant for the MiniBooNE~\cite{Mini}, T2K~\cite{T2K}, and MINER$\nu$A~\cite{MINERVA} experiments.  Their incoming neutrino fluxes are characterized by average energies ranging from  $700~\mathrm{MeV}$ for T2K up to $6 ~\mathrm{GeV}$ of the medium-energy NuMI beam in MINER$\nu$A. Therefore, the cross section receives contributions from the high momentum region of the phase space, where a proper treatment of relativistic effects become relevant. We account for the latter by evaluating the GFMC electroweak responses in the ANB frame and boosting them back to the LAB fram. As argued above, since the ANB frame minimizes relativistic effects, we find that applying the two-fragment model brings about minimal differences. 

\begin{figure*}
    \centering
    \includegraphics[width=0.32\textwidth]{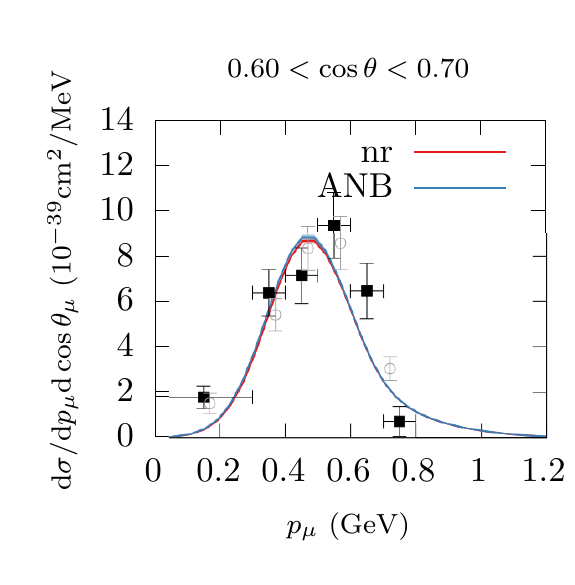}
    \includegraphics[width=0.32\textwidth]{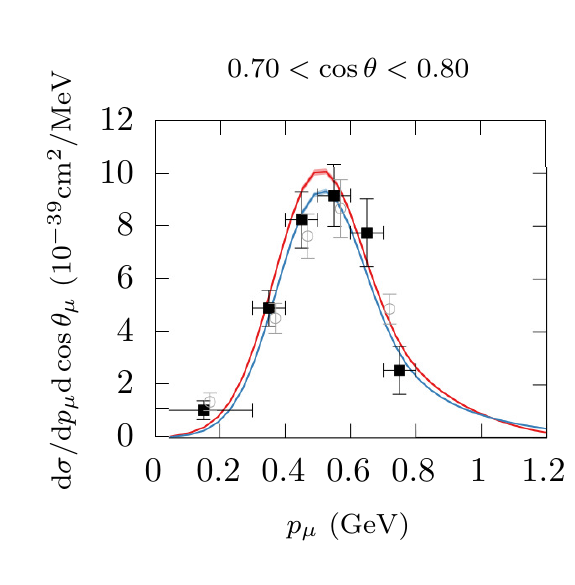}
    \includegraphics[width=0.32\textwidth]{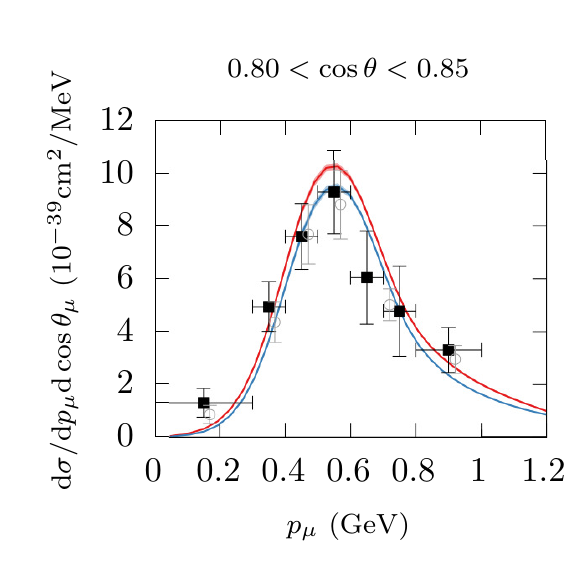} \\
    \includegraphics[width=0.32\textwidth]{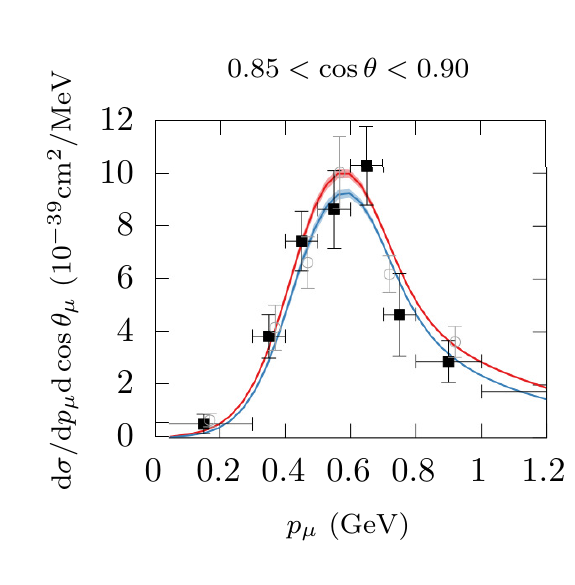}
    \includegraphics[width=0.32\textwidth]{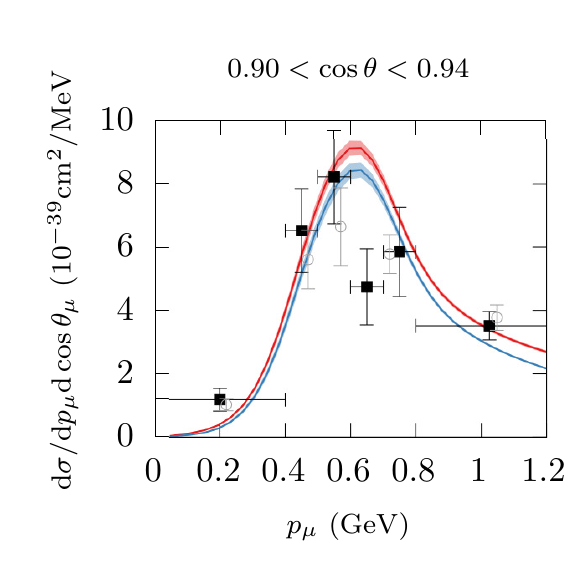}
    \includegraphics[width=0.32\textwidth]{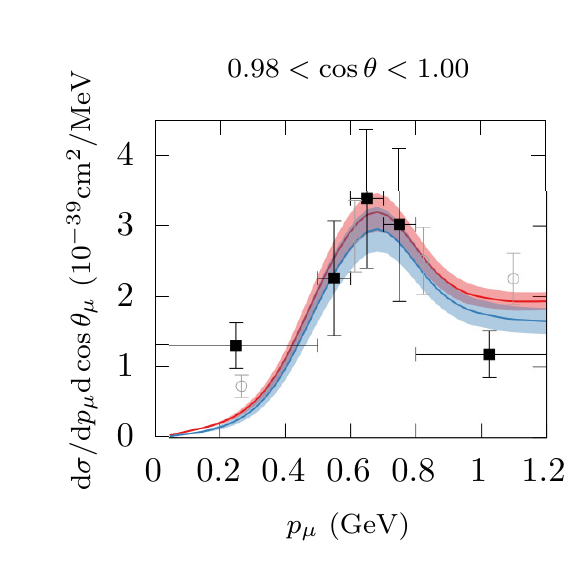}
    \caption{T2K flux folded GFMC results, nonrelativistic (nr), and in the ANB, both including one- and two-body current contributions. The black data points are from Ref.~\cite{T2K:Inclnumu}, while the data from the analysis reported in Ref.~\cite{T2K:nuanu} is shown by the gray data points.}
    \label{fig:T2K_fluxfolded_12b}
\end{figure*}

\subsection{MiniBooNE}
Our theoretical calculations for the flux averaged double differential cross section for the MiniBooNE kinematics are shown in Fig.~\ref{fig:MB_fluxfolded_12b}. Both the nonrelativistic and ANB results include one- and two-body current contributions. 
The black squares correspond to the `CCQE-like' data reported in Ref.~\cite{MB:QECS}, whose extraction from experimental measurements entails some model dependence~\cite{TENSIONS19}. In particular, an irreducible 'non-CCQE' background, mainly consisting of the production of a single $\pi^+$ which is either absorbed or remains otherwise undetected~\cite{Alvarez-Ruso:2017oui,NuSTEC:2020nsl,Ruso:2022qes}, is estimated using the NUANCE generator~\cite{NUANCE}, and subtracted from the data. 
This background is partly constrained by their own measurement~\cite{MB:pion}, but inconsistencies in the description of the MiniBooNE $\pi^+$ production data and data from T2K~\cite{T2KCC1PIH2O} and MINER$\nu$A~\cite{MINERvA:CC1PI} have been pointed out~\cite{Sobczyk15, TENSIONS19, Nikolakopoulos18a, Gonzalez-Jimenez18}.
Hence, to better gauge the uncertainties associated with this procedure, it is best practice to add this background back to the data points; we show the resulting distribution in Fig.~\ref{fig:MB_fluxfolded_12b} as empty circles.
Finally, one should keep in mind that the MiniBooNE collaboration reports an overall 10\% normalization error which is not taken into account in the error-bars. 

\begin{figure*}
    \includegraphics[width=\textwidth]{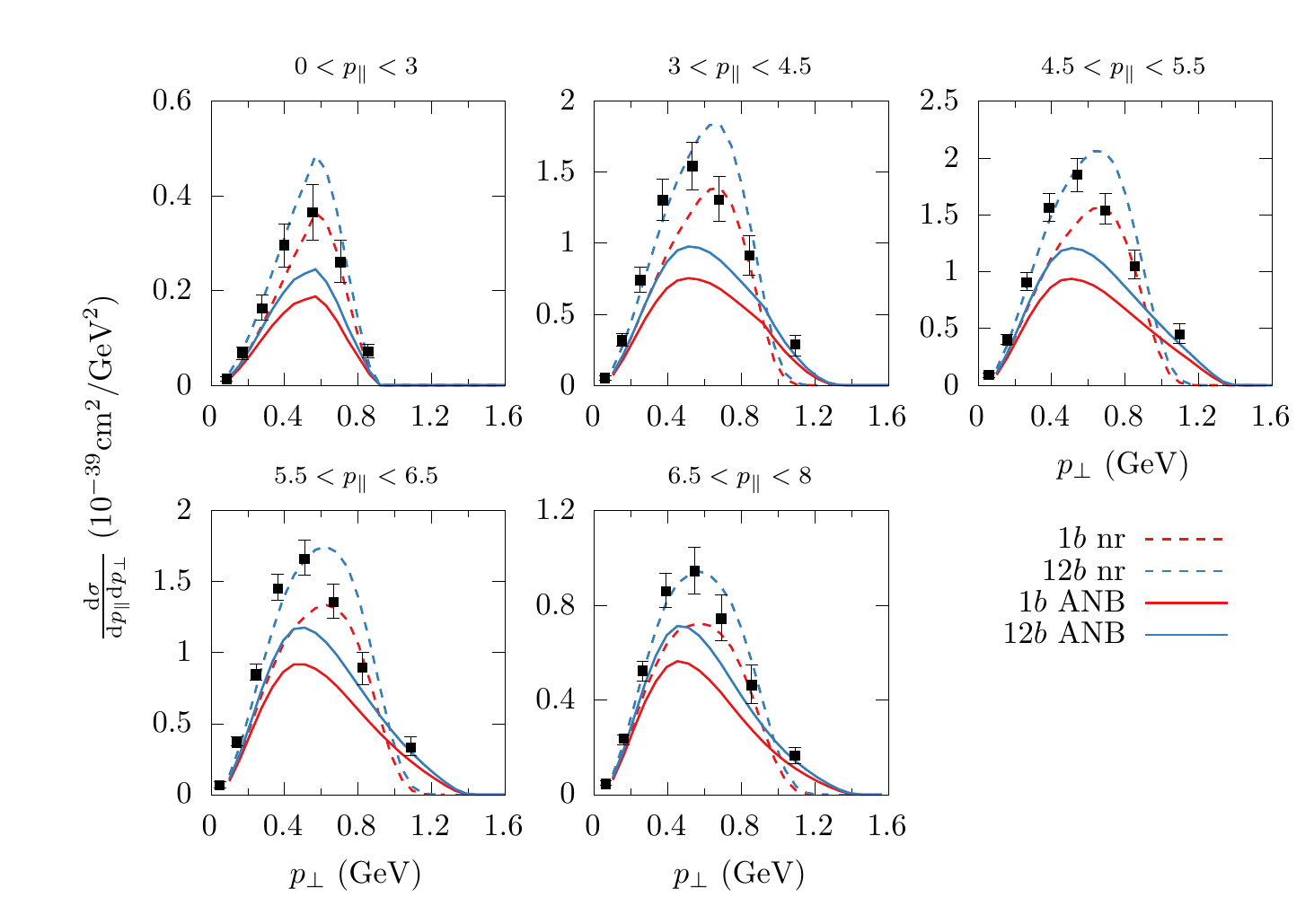}
    \caption{Flux-averaged cross sections for MINER$\nu$A in terms of $p_\perp$ in different bins of $p_\parallel$. Nonrelativistic results are shown by dashed lines, while the ones that include relativistic corrections are shown by solid lines. The experimental data are from Ref.~\cite{MINERvA:2023kuz}}
    \label{fig:Minerva}
\end{figure*}

The effect of the relativistic corrections implemented through the ANB response is a reduction of the peak strength with a redistribution towards larger values of $T_\mu$.
It is interesting to note that the calculations tend to saturate the data at small $T_\mu$, while leaving space at large $T_\mu$, as has been previously pointed out in Refs.~\cite{Jachowicz:EPST,Lovato:PRX}.
The present calculations use a dipole parametrization of the axial form factor with a cut-off $M_A = 1\sim\mathrm{GeV}$. However, recent Lattice-QCD calculations suggest a significantly larger axial form factor at $Q^2=\mathbf{q}^2-\omega^2\sim 1$ GeV$^2$~\cite{RQCD:2019jai,Jang:2019vkm,Park:2021ypf}. Including an axial form factors consistent with these Lattice-QCD results in GFMC and spectral-function calculations~\cite{Simons:2022ltq} increases the inclusive cross sections at high-$T_\mu$, compared to a dipole with $M_A = 1\sim\mathrm{GeV}$. This enhancement is consistent with earlier works~\cite{MB:QECS} based on simplified models of nuclear dynamics. On the other hand, a number of neutrino event generators that use a dipole form with $M_A\approx 1\sim\mathrm{GeV}$ provide a reasonable description of the MiniBooNE data, once the model-dependent background is added~\cite{TENSIONS19}. Notably, in this latter comparison, the data points seem to be shifted to smaller $T_\mu$.

The relativistic corrections computed in this work are critical to perform meaningful comparisons between GFMC calculations and MiniBooNE data~\cite{Lovato:PRX}. In particular, including relativistic effects is critical to test different parameterizations of the axial form factor. However, the uncertainties in the MiniBooNE analysis hamper a firm conclusions in a theory-data comparison. In view of the statistical significance of the MiniBooNE dataset, the unresolved tensions with other experiments, and the possible importance for informing modeling in the SBN program at Fermilab, a reanalysis of the MiniBooNE dataset(s) would be immensely beneficial~\cite{TENSIONS19}.

\subsection{T2K}
Fig.~\ref{fig:T2K_fluxfolded_12b} displays our results for the T2K experiment using the flux tabulated in Ref.~\cite{T2K:flux2016}. The GFMC calculations again include one and two-body terms in the charged-current operator. The two sets of data correspond to the original analysis of Ref.~\cite{T2K:nuanu} and the more recent one reported in Ref.~\cite{T2K:Inclnumu}. As expected, the difference between the calculations carried out in the ANB frame and the nonrelativistic ones is much smaller than for MiniBooNE, owing to the lower average energy in the T2K flux. Experimental data are well reproduced by the one plus two-body current theoretical results, leaving little room for higher-energy reaction mechanisms. In this regard, for this kinematics using the Lattice-QCD axial form factor brings about minor differences compared to the dipole one with $M_A = 1\sim\mathrm{GeV}$~ Ref.~\cite{Simons:2022ltq}

\subsection{MINER$\nu$A}

\begin{figure*}
    \includegraphics[width=\textwidth]{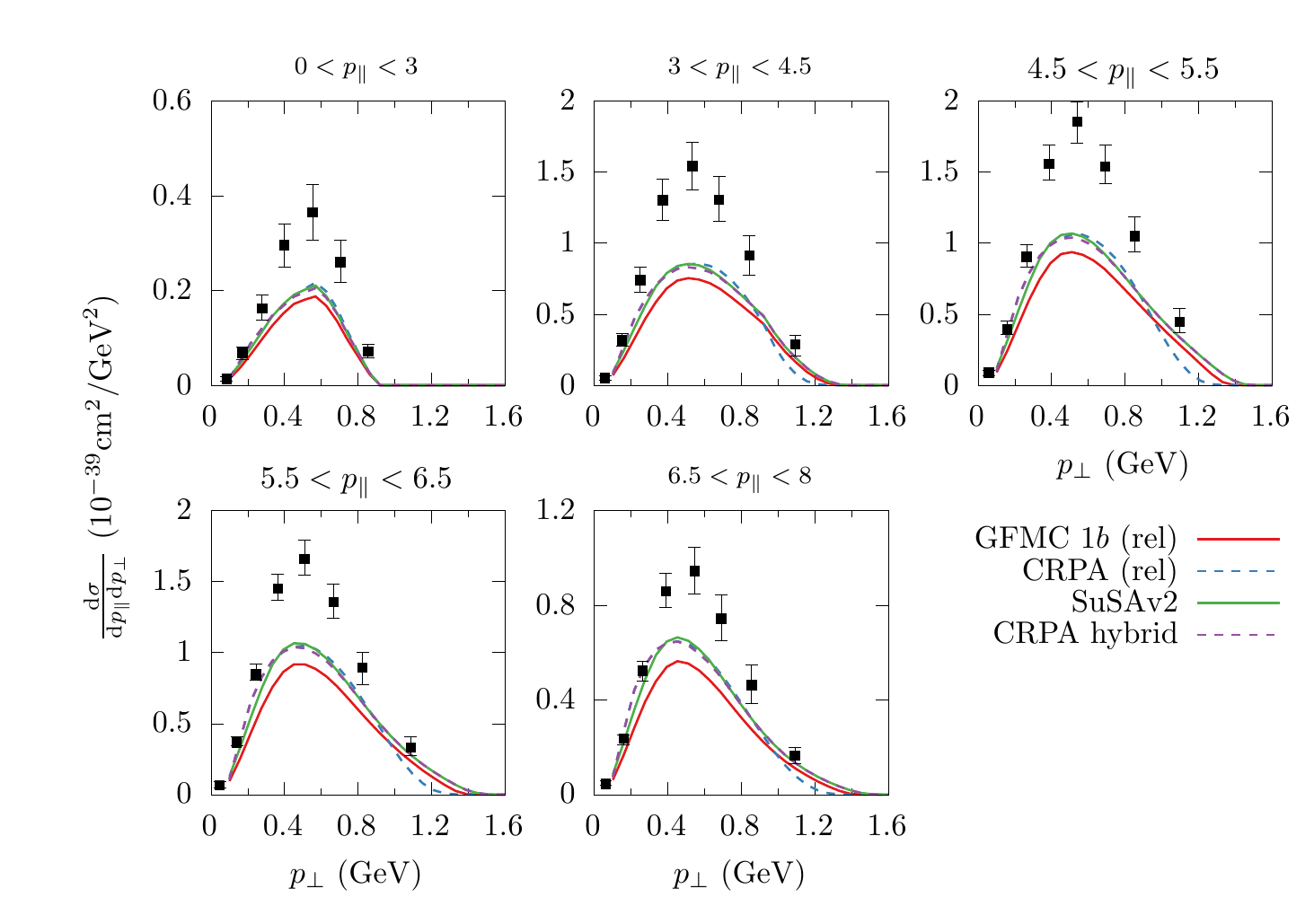}
    \caption{Flux-averaged cross sections for MINER$\nu$A in terms of $p_\perp$ in different bins of $p_\parallel$. Theoretical calculations only retain one-body current contributions and are carried out within microscopic (GFMC) and mean-field (CRPA, SuSAv2, and CRPA hybrid) approaches to nuclear dynamics.}
    \label{fig:Minerva_1bmodels}
\end{figure*}

In Figs.~\ref{fig:Minerva} and~\ref{fig:Minerva_1bmodels} we show the MINER$\nu$A CC cross section results 
as a function of longitudinal and transverse muon momentum. 
These are defined as
\begin{equation}
    p_{\parallel} =  \lvert \mathbf{p}_\mu \rvert \cos\theta_\mu 
\end{equation}
and
\begin{equation}
    p_{\perp} = \lvert \mathbf{p}_\mu \rvert \sin\theta_\mu = \sqrt{\mathbf{p}_\mu ^2 - p_\parallel^2 }, 
\end{equation}
respectively, $\theta_\mu$ being the scattering angle with respect to the beam. The differential cross section is then
\begin{equation}
    \frac{d^2\sigma}{dp_\perp dp_\parallel} = \frac{p_\perp}{\lvert \mathbf{p}_\mu \rvert E_\mu}  \frac{d^2\sigma}{dE_\mu d\cos\theta_\mu}.
\end{equation}
The data is obtained by exposure to the medium-energy NuMI beam; we use the flux of Ref.~\cite{NUMI:ME2019} and compare with the cross section data of Ref.~\cite{MINERvA:2023kuz}.
Both the data and calculations include kinematics cuts in the scattering angle $\theta_\mu < 17^{\circ}$ and the muon momentum $2~\mathrm{GeV} < \lvert \mathbf{p}_\mu\rvert < 20~\mathrm{GeV}$.
Additionally we restrict all calculations to $ E_\nu < 20~\mathrm{GeV}$ and momentum transfer $|\qv| < 2~\mathrm{GeV}$.

The comparison of the purely nonrelativistic and the ANB results are shown in Fig.~\ref{fig:Minerva}.
The inclusion of relativistic effects reduces the cross section by almost a factor of two for low-$p_\parallel$, with the difference in magnitude around the peak decreasing for larger $p_\parallel$.
We note that the momentum transfer is limited as $q > p_{\perp}$, and that bins at small $p_\parallel$ generally allow for higher energy, and hence larger $q$ contributions at small $p_{\perp}$, which explains this behavior.
The appearance of the high-$p_\perp$ (i.e. high-$q$) tails can be understood by the narrowing of the response in terms of the energy transfer compared to the nonrelativistic results --- see Fig.~\ref{fig:C12_scaling_r_nr} --- that redistributes strength into the available phase space at large-$q$.

As calculations for MINER$\nu$A include large $q$, and the effect of the relativistic corrections is significant, a consistency check is in order.
For this reason, in Fig.~\ref{fig:Minerva_1bmodels} we compare the GFMC calculations that only include one-body currents to other approaches, based on a mean-field approximation of nuclear dynamics. Specifically, the CRPA calculations~\cite{Pandey:2016, Jachowicz02, Jachowicz:EPST}, include the relativistic correction in the nucleon energy discussed above~\cite{Amaro:2005}. The SuSAv2 results are based on the relativistic scaling formalism --- for additional details see Ref.~\cite{SuSav2}. Finally the ``Hybrid'' CRPA calculations introduce a blending of the nuclear responses with SuSAv2 responses in the region $500 \lesssim q \lesssim 700$ MeV. Above this region, the results are purely SuSAv2, and below they are purely CRPA. In the region in between, the SuSAv2 and CRPA results are practically identical~\cite{Dolan:2022CRPA}.
We find the different theoretical calculations to be in reasonably good agreement. The strength of the GFMC one-body contribution is quenched by a $\sim 10\%$ with respect to the other curves, and the tail of the CRPA calculations at high-$p_T$ drop faster. This might be ascribed to the broadening of the CRPA responses at high-$q$ compared to the SuSAv2 results, see e.g. discussion in Ref.~\cite{Dolan:2022CRPA}.

Finally, we comment on the fact that the nonrelativistic calculations seems to be in better agreement with experimental data than the ones in which relativistic effects are accounted for. However, given the energy distribution of the medium-energy NuMI beam of the MINER$\nu$A experiment, contributions beyond quasi-elastic scattering are expected to be significant, even if the experimental analysis rejects events with mesons visible in the detector. In particular, there are instances in which pions produced in the interactions vertex are is either absorbed or remain undetected. Hence, theoretical calculations that do not include pion-production mechanisms should remain below experimental data. This is indeed the case when relativistic effects are accounted for, while neglecting them yields unphysically large cross sections.

\section{Conclusions}
\label{sec:conclusions}
One of the main sources of systematic uncertainties in neutrino-oscillation experiments comes from the limited accuracy in the prediction of neutrino-nucleus cross sections. Using sophisticated QMC techniques, in particular the GFMC approach, has proven to be successful in describing electroweak interactions for low to moderate momentum transfer in the quasielastic region, where the dominant reaction mechanisms are single- and multi-nucleon knockout. The main shortcomings of the GFMC lie in its nonrelativistic nature and being limited to inclusive predictions. The evaluation of the GFMC nuclear electromagnetic responses in a reference frame that minimizes relativistic effects, namely, the ANB frame, has been discussed in Ref.~\cite{Rocco:2018}. This strategy appeared to be successful in accounting for relativistic corrections in the kinematics for nuclei with $A=3,4$ nucleons. In this work, we extend the approach to $^{12}$C and consider charged-current interactions, in which the axial component violates current conservation. Differently from Refs.~\cite{Efros:2005uk,Efros:2009qp,Rocco:2018}, to properly transform the axial term, we adopt the general expression of the Lorentz-boost to connect the responses evaluated in different frames. In addition to working in the ANB frame, we consider different strategies to incorporate relativistic effects in the kinematics. Following Refs.~\cite{Efros:2005uk,Efros:2009qp,Rocco:2018}, we implement the two-fragment model in which the kinematic inputs of the nonrelativistic dynamical calculation are obtained in a relativistically correct fashion. An alternative approach based based on shifting the outgoing nucleon energy when solving the Shr\"{o}dinger equation in the CRPA approach is also considered. We argue that these different methods produce similar corrections in the GFMC and CRPA calculations of the electroweak response at $q=700$ MeV. 

We compute the CC flux-averaged neutrino cross-section within GFMC including one- and two-body current operators and compare it with experimental data from the T2K, MiniBooNE, and MINER$\nu$A collaborations. Since the average neutrino energy for the T2K beam is around $700$ MeV, we find that relativistic corrections are in general very small, and only visible for some values of the scattering angle. On the other hand, while the average neutrino energy of the MiniBooNE experiment is also of the order of $700$ MeV, the tails of the flux extend up to $3$ GeV. In this case, we observe that the inclusion of relativistic effects in the kinematics yields a visible reduction in the strength at the quasielastic peak for all the scattering angles considered. Hence, accounting for relativistic effects is critical for testing different parametrizations of the nucleon axial form factor, including those recently obtained within Lattice-QCD~\cite{RQCD:2019jai,Jang:2019vkm,Park:2021ypf}.

Finally, we gauge relativistic effects in GFMC calculations in the extreme case of MINER$\nu$A kinematics, where the medium-energy NUMI beam peaks around $6$ GeV. Including relativistic corrections has a dramatic effect, yielding a reduction of the strength up to $50\%$ compared to nonrelativistic calculations. Despite working in the ANB frame has proven effective in accounting for relativistic corrections, we do not expect the GFMC to be applicable in this high energy regime. For this reason, we compare the results obtained from the GFMC calculations in the ANB frame with other approaches allowing for a fully relativistic treatment of the kinematics, such as SuSAv2. The good agreement between the relativistically-corrected GFMC cross sections and SuSaA2 results corroborates the validity of the procedure we employ to include relativistic effects in the GFMC.

\acknowledgments{
This manuscript has been authored by Fermi Research Alliance, LLC under Contract No. DE-AC02-07CH11359 with the U.S. Department of Energy, Office of Science, Office of High Energy Physics and by the NeuCol SciDAC-5 program (N.R. and A.L.). The present research is also supported by the U.S. Department of Energy, Office of Science, Office of Nuclear Physics, under contracts DE-AC02-06CH11357 (A.L.), by the NUCLEI SciDAC-5 program (A.L.), the DOE Early Career Research Program (A.L.), and Argonne LDRD awards (A.L.). Quantum Monte Carlo calculations were performed on the parallel computers of the Laboratory Computing Resource Center, Argonne National Laboratory, the computers of the Argonne Leadership Computing Facility via INCITE and ALCC grants.}

\bibliographystyle{apsrev4-1.bst}
\bibliography{bibliography}

\end{document}